# RELATIVISTIC TREATMENT OF THE HELLMANN-GENERALIZED MORSE POTENTIAL

P. O. Okoi[1], C. O. Edet[2] and T. O. Magu[3&4]

[1]Department of Physics, University of Calabar, Nigeria

[2]Theoretical Physics Group, Department of Physics, University of Port Harcourt, Choba, Nigeria

[3]Department of Pure and Applied Chemistry, University of Calabar, Calabar, Nigeria

[4]CAS Key Laboratory of Green Printing, Institute of Chemistry, University of Chinese Academy of Sciences, 100190 Beijing, P. R. China.

## Abstract

We solve the relativistic equations(Klein-Gordon and Dirac equation) via the conventional Nikiforov-Uvarov method. In order to overcome the centrifugal barrier, we employed the well-known Greene and Aldrich approximation scheme. The corresponding normalized eigenfunctions was also obtained in each case. It was shown that in the non-relativistic limits, both energy equations obtained by solving Klein-Gordon and Dirac equations, and wavefunctions reduced to the non-relativisitc energy Equation. The bound state energy eigenvalues for $N_2$, CO, NO, CH and HCl diatomic molecules were computed for various vibrational and rotational quantum numbers. It was found that our results agree with those in literature.

**Keywords**; Hellmann-generalized Morse potential; Dirac Equation; Klein-Gordon equation; Nikiforov–Uvarov method; Schrodinger equation.

## 1. INTRODUCTION

Many contemporary problems in atomic, nuclear, solid state, rotational and vibrational spectroscopy involve obtaining the exact or approximate solutions of the relativistic and non-relativistic wave equation describing the dynamics of the quantum system under consideration. This is so because the wavefunction contains all the necessary information needed to describe the behaviour of the system. Recently, many authors have obtained both the exact ($\ell = 0$) and approximate ($\ell \neq 0$) solutions of some exponential type potentials which include inverted generalized hyperbolic potential [1], Hulthen potential [2-4],Rosen-Morse potential[5] Poschl-Teller potential [6,7], Eckart potential [8], modified Hylleraas potential [9], generalized Morse potential [10], Woods-Saxon potential [11], Manning-Rosen potential [12,13] among others. This was achieved by employing appropriate approximation schemes Pekeris [14], Greene and Aldrich [15] and even more recently improved ones [2,16-18] to accommodate large values of the screening parameters.

Some of the techniques devised to investigate these potentials are; Asymptotic Iteration Method (AIM) [5,13,19], Nikiforov-Uvarov (NU) method [20,21], shape invariant supersymmetry (SUSYQM) [22], Modified factorization method (MFM) [23], Formula method [24], Exact quantization rule [25-27], Factorization method [28] and others.

**E-mail for correspondence:** collinsokonedet@gmail.com

The Klein-Gordon equation containing a four vector linear momentum operator and a rest mass requires introducing the four vector potential $V(r)$ and a space time scalar potential $S(r)$. With the configuration $S(r) = V(r)$ or $S(r) = -V(r)$, it has been shown extensively in literature that the Klein-Gordon equation and Dirac Equation share the same energy spectrum [29]. While for $S(r) = V(r) = 2V(r)$, gives non-relativistic limits of the equation conforming exactly to that of the Schrodinger equation [30-32].

The Hellmann potential [33-36] is a short range potential. It is a combination of Yukawa and Coulomb potentials. This potential have been investigated extensively by various authors. For instance, [36] used the shape invariant supersymmetric approach to solve analytically the three-dimensional Schrödinger equation with this potential by applying the well-known Greene and Aldrich approximation[37] scheme to deal with the centrifugal barrier. [38] obtained approximate Eigensolutions of the Duffin-Kemmer-Petiau and Klein-Gordon equations with Hellmann potential. [39] solved the Schrodinger Equation with Hellmann potential using the parametric Nikiforov-Uvarov method. This potenial finds more application in the field of atomic and condensed matter physic, electron-core [40,41], electron-ion [42] inner-shell ionization problem, alkali hydride molecules and solid state physics [43,44].

In addition, the Deng-Fan Oscillator Potential [45,46] is applied in molecular physics, atomic physics and quantum chemistry. It is used to describe the interactions of molecular structure in quantum mechanics for diatomic molecules. This potential is the generalized Morse potential [47, 48]. It has physical boundary conditions at r = 0 and ∞. It has the correct asymptotic behaviour as the inter-nuclear distance approaches 0 [47, 48]. and it has been used extensively to describe diatomic molecular energy spectra and electromagnetic transitions [49]. The bound state solutions of the relativistic and non-relativistic wave equations have been studied by several authors with this potential [27, 50–52]

Also, the authors of Ref. [53] proposed the Hellmann-generalized Morse potential model and solved the Schrodinger equation with it. Motivated by the success of these investigation, we attempt to solve the Klein-Gordon and Dirac Equations with this potential using Nikiforov-Uvarov method. We'll also apply the Non- Relativistic Energy eqation obtained to study some selected diatomic molecules with the aid of some spectroscopic parameters. The potential is of the form[53];

$$V(r) = D_e \left[ 1 - \frac{a + be^{-\alpha r}}{rD_e} - 2\left(\frac{e^{\alpha r_e} - 1}{e^{\alpha r} - 1}\right) + \left(\frac{e^{\alpha r_e} - 1}{e^{\alpha r} - 1}\right)^2 \right] \qquad (1)$$

The work is organised in the following way; in the next section, we review the NU method. The eigensolutions of Klein-Gordon equation and Dirac equation with the Hellmann- generalized Morse potential and their non-relativistic limits conforming to the spectrum of the Schrodinger equation are obtained in section 3, while section 4 presents the numerical results with plots for the energy spectrum. Finally, we give our concluding remarks of the findings in section 5.

## 2. REVIEW OF NIKIFOROV-UVAROV METHOD

The Nikiforov-Uvarov (NU) method is based on solving the hypergeometric-type second-order differential equations by means of the special orthogonal functions . The main equation which is closely associated with the method is given in the following form

$$\psi''(s) + \frac{\tilde{\tau}(s)}{\sigma(s)}\psi'(s) + \frac{\tilde{\sigma}(s)}{\sigma^2(s)}\psi(s) = 0 \qquad (2)$$

Where $\sigma(s)$ an $\tilde{\sigma}(s)$ are polynomials at most second-degree, $\tilde{\tau}(s)$ is a first-degree polynomial and $\psi(s)$ is a function of the hypergeometric-type.

The exact solution of Eq. (2) can be obtained by using the transformation

$$\psi(s) = \phi(s) y(s) \qquad (3)$$

This transformation reduces Eq. (2) into a hypergeometric-type equation of the form

$$\sigma(s) y''(s) + \tau(s) y'(s) + \lambda y(s) = 0 \qquad (4)$$

The function $\phi(s)$ can be defined as the logarithm derivative

$$\frac{\phi'(s)}{\phi(s)} = \frac{\pi(s)}{\sigma(s)} \qquad (5)$$

where $\pi(s) = \frac{1}{2}\left[\tau(s) - \tilde{\tau}(s)\right] \qquad (5a)$

with $\pi(s)$ being at most a first-degree polynomial. The second $\psi(s)$ being $\chi_n(s)$ in Eq. (3), is the hypergeometric function with its polynomial solution given by Rodrigues relatio

$$\chi_n^{(n)}(s) = \frac{B_n}{\rho(s)} \frac{d^n}{ds^n}\left[\sigma^n \rho(s)\right] \qquad (6)$$

Here, $B_n$ is the normalization constant and $\rho(s)$ is the weight function which must satisfy the condition

$$(\sigma(s)\rho(s))' = \sigma(s)\tau(s) \qquad (7)$$

$$\tau(s) = \tilde{\tau}(s) + 2\pi(s) \qquad (8)$$

It should be noted that the derivative of $\tau(s)$ with respect to $s$ should be negative. The eigenfunctions and eigenvalues can be obtained using the definition of the following function $\pi(s)$ and parameter $\lambda$, respectively:

$$\pi(s) = \frac{\sigma'(s) - \tilde{\tau}(s)}{2} \pm \sqrt{\left(\frac{\sigma'(s) - \tilde{\tau}(s)}{2}\right)^2 - \tilde{\sigma}(s) + k\sigma(s)} \qquad (9)$$

where $k = \lambda - \pi'(s)$ \hfill (10)

The value of $k$ can be obtained by setting the discriminant of the square root in Eq. (9) equal to zero. As such, the new eigenvalue equation can be given as

$$\lambda_n = -n\tau'(s) - \frac{n(n-1)}{2}\sigma''(s), \quad n = 0,1,2,.... \qquad (11)$$

### 3.1 Bound state solution of the Klein-Gordon equation

The Klein-Gordon equation for a spinless particle for $\hbar = c = 1$ in D-dimensions is given as[62]

$$\left[-\nabla^2 + (M + S(r))^2 + \frac{(D+2l-1)(D+2l-3)}{4r^2}\right]\psi(r,\theta,\varphi) = [E - V(r)]^2 \psi(r,\theta,\varphi) \qquad (12)$$

Where $\nabla^2$ is the Laplacian, $M$ is the reduced mass, $E_{nl}$ is the energy spectrum, and $n$ and $l$ are the radial and orbital angular momentum quantum numbers respectively (or vibration-rotation quantum number in quantum chemistry). It is a common practise that for the wavefunction to satisfy the boundary conditions it can be rewritten as $\psi(r,\theta,\varphi) = R_{nl}/r \, Y_{lm}(\theta,\varphi)$. However, the spherical harmonics $Y_{lm}(\theta,\varphi)$ is already known in literature [54] and the angular component of the wavefunction could be separated leaving only the radial part.

So the radial part of the equation becomes

$$\frac{d^2 R_{nl}}{dr^2} + \left[\left(E_{nl}^2 - M^2\right) + V^2(r) - S^2(r) - 2(EV(r) + MS(r)) - \frac{(D+2l-1)(D+2l-3)}{4r^2}\right]R_{nl} = 0 \qquad (13)$$

Thus, for equal vector and scalar potentials $V(r) = S(r) = 2V(r)$ Eq. (14) becomes

$$\frac{d^2 R_{nl}}{dr^2} + \left[\left(E_{nl}^2 - M^2\right) - V(r)(E_{nl} + M) - \frac{(D+2l-1)(D+2l-3)}{4r^2}\right]R_{nl} = 0 \qquad (14)$$

The Greene and Aldrich approximation scheme [37] for the centrifugal term of the Klein-Gordon equation is given as

$$\frac{1}{r^2} \approx \frac{\alpha^2}{\left(1 - e^{-\alpha r}\right)^2} \qquad (15)$$

Using Eq. (1) with $q = e^{\alpha r_e} - 1$, then, Eq. (14) becomes

$$\frac{d^2 U_{nl}}{dr^2} + \left[\begin{array}{c}\left(E_{nl}^2 - M^2\right) - \left(D_e - \dfrac{a}{r} + \dfrac{be^{-\alpha r}}{r} - \dfrac{2D_e q e^{-\alpha r}}{1-e^{-\alpha r}} + \dfrac{D_e q^2 e^{-2\alpha r}}{\left(1-e^{-\alpha r}\right)^2}\right)\left(E_{nl} + M\right) \\ -\dfrac{(D+2l-1)(D+2l-3)}{4r^2}\end{array}\right] U_{nl} = 0 \quad (16)$$

using $s = e^{-\alpha r}$ so as to enable us apply the NU method and substituting Eq. (15) into (16), we have

$$\frac{d^2 U_{nl}}{dr^2} + \frac{(1-s)}{s(1-s)}\frac{dU_{nl}}{dr} + \frac{1}{s^2(1-s)^2}\left[\begin{array}{c}\dfrac{\left(E_{nl}^2 - M^2\right) - D_e\left(E_{nl} + M\right)}{\alpha^2}(1-s)^2 + \dfrac{a}{\alpha}\left(E_{nl} + M\right)(1-s) \\ -\dfrac{b}{\alpha}\left(E_{nl} + M\right)s(1-s) + \dfrac{2D_e q\left(E_{nl} + M\right)}{\alpha^2}s(1-s) \\ -\dfrac{D_e q^2\left(E_{nl} + M\right)}{\alpha^2}s^2 - \dfrac{(D+2l-1)(D+2l-3)}{4}\end{array}\right] U_{nl} = 0 \quad (17)$$

Eq. (17) can be simplified further by using the following dimensionless ansatz

$$-\varepsilon_{nl} = \frac{\left(E_{nl}^2 - M^2\right) - D_e\left(E_{nl} + M\right)}{\alpha^2},\ \beta = \frac{a}{\alpha}\left(E_{nl} + M\right),\ \eta = \frac{b}{\alpha}\left(E_{nl} + M\right),\ \chi = \frac{2D_e q\left(E_{nl} + M\right)}{\alpha^2},$$

$$\varphi = \frac{D_e q^2\left(E_{nl} + M\right)}{\alpha^2} \text{ and } \gamma = \frac{(D+2l-1)(D+2l-3)}{4} \quad (18)$$

Substituting the ansatz and simplifying, we obtain

$$\frac{d^2 U_{nl}}{dr^2} + \frac{(1-s)}{s(1-s)}\frac{dU_{nl}}{dr} + \frac{1}{s^2(1-s)^2}\left[\begin{array}{c}-\left(\varepsilon_{nl} - \eta + \chi + \varphi\right)s^2 + \left(2\varepsilon_{nl} - \beta - \eta + \chi\right)s \\ -\left(\varepsilon_{nl} - \beta + \gamma\right)\end{array}\right] U_{nl} = 0 \quad (19)$$

Comparing Eq. (19) with Eq. (2), we obtain the following polynomials;

$$\sigma(s) = s(1-s),\ \tilde{\tau}(s) = 1 - s$$
$$\tilde{\sigma}(s) = -\left(\varepsilon_{nl} - \eta + \chi + \varphi\right)s^2 + \left(2\varepsilon_{nl} - \beta - \eta + \chi\right)s - \left(\varepsilon_{nl} - \beta + \gamma\right) \quad (20)$$

Substituting these polynomials into Eq. (9), we get $\pi(s)$ to be

$$\pi(s) = \frac{-s}{2} \pm \sqrt{\left(\frac{1}{4} + \varepsilon_{nl} - \eta + \chi + \varphi - k\right)s^2 + \left(k - \left(2\varepsilon_{nl} - \beta - \eta + \chi\right)\right)s + \varepsilon_{nl} - \beta + \gamma} \quad (21)$$

To find the value for $k$, the discrimant of the expression in the square root in Eq. (21) must vanish. So we obtain

$$k = -\left(-\beta + \eta - \chi + 2\gamma\right) \pm 2\left(\varepsilon_{nl} - \beta + \gamma\right)\sqrt{\frac{1}{4} + \varphi + \gamma} \quad (22)$$

Substituting $k_-$ of Eq. (22) into Eq. (21) yields

$$\pi(s) = \frac{-s}{2} \pm \left[\left(\sqrt{\varepsilon_{nl} - \beta + \gamma} + \sqrt{\frac{1}{4} + \varphi + \gamma}\right)s - \sqrt{\varepsilon_{nl} - \beta + \gamma}\right] \quad (23)$$

From Eq. (8), the condition for the bound state solution of the NU equation is satisfied thus

$$\tau'(s) = -2\left(1 + \sqrt{\varepsilon_{nl} - \beta + \gamma} + \sqrt{\frac{1}{4} + \varphi + \gamma}\right) < 0 \quad (24)$$

However, the parameter $\lambda$ is obtained from Eq. (10) to be

$$\lambda = -(-\beta + \eta - \chi + 2\gamma) - 2\sqrt{\varepsilon_{nl} - \beta + \gamma}\sqrt{\frac{1}{4} + \varphi + \gamma} - \frac{1}{2} - \sqrt{\varepsilon_{nl} - \beta + \gamma} - \sqrt{\frac{1}{4} + \varphi + \gamma} \quad (25)$$

And employing Eq. (11), we obtain

$$\lambda_n = n^2 + n + 2n\sqrt{\varepsilon_{nl} - \beta + \gamma} + 2n\sqrt{\frac{1}{4} + \varphi + \gamma} \quad (26)$$

The condition for obtaining bound state energy eigenvalue equation, requires that $\lambda = \lambda_n$, so equating this, and simplifying yields

$$\varepsilon_{nl} = \beta - \gamma + \frac{1}{4}\left[\frac{\left(n + \frac{1}{2} + \sqrt{\frac{1}{4} + \varphi + \gamma}\right)^2 - \beta + \eta - \chi + \gamma - \varphi}{\left(n + \frac{1}{2} + \sqrt{\frac{1}{4} + \varphi + \gamma}\right)}\right]^2 \quad (27)$$

Where upon substituting the ansatz of Eq. (18) into (27) gives the energy eigenvlaue equation

$$E_{nl}^2 - M^2 = (D_e - a\alpha)(E_{nl} + M) + \alpha^2 \Lambda - \frac{1}{4}\left[\frac{\alpha\left(n + \frac{1}{2} + \delta\right)^2 - (E_{nl} + M)\left(a - b + \frac{2D_e}{\alpha}\left(e^{\alpha r_e} - 1\right) + \frac{D_e}{\alpha}\left(e^{\alpha r_e} - 1\right)^2\right) + \alpha\Lambda}{\left(n + \frac{1}{2} + \delta\right)}\right]^2 \quad (28)$$

Where
$$\begin{cases} \delta = \sqrt{\dfrac{1}{4}+\dfrac{D_e}{\alpha^2}(E_{nl}+M)(e^{\alpha r_e}-1)^2+\Lambda} \\ \Lambda = \dfrac{(D+2l-1)(D+2l-3)}{4} \end{cases} \qquad (29)$$

## 3.2 Non-relativistic Limit

In this section, we consider the non-relativistic limit of Eq. (28). Considering a transformation of the form: $M + E_{n\ell} \to \dfrac{2\mu}{\hbar^2}$ and $M - E_{n\ell} \to -E_{n\ell}$ and substitute it into Eqn. (28), we have the nonrelativistic energy equation as

$$E_{n\ell} = D_e - a\alpha + \frac{\hbar^2\alpha^2 \ell(\ell+1)}{2\mu} - \frac{\hbar^2\alpha^2}{8\mu}\left[\frac{\left(n+\frac{1}{2}+\sqrt{\frac{1}{4}+\ell(\ell+1)+\frac{2\mu D_e q^2}{\hbar^2\alpha^2}}\right)^2 + \frac{2\mu}{\hbar^2}\left(\frac{b}{\alpha}-\frac{2D_e q}{\alpha^2}-\frac{a}{\alpha}+\frac{D_e q^2}{\alpha^2}\right)+\ell(\ell+1)}{\left(n+\frac{1}{2}+\sqrt{\frac{1}{4}+\ell(\ell+1)+\frac{2\mu D_e q^2}{\hbar^2\alpha^2}}\right)}\right]^2 \qquad (30)$$

To obtain the corresponding wavefunction, we consider Eq. (5), and upon substituting Eq. (18) and (20) and integrating, we get

$$\phi(s) = s^{\sqrt{\varepsilon_{nl}-\beta+\gamma}}(1-s)^{\frac{1}{2}+\sqrt{\frac{1}{4}+\varphi+\gamma}} \qquad (31)$$

To get the hypergeometric function considering Eq. (3), we first calculate the weight function following Eq. (7) as

$$\rho(s) = s^{2\sqrt{\varepsilon_{nl}-\beta+\gamma}}(1-s)^{2\sqrt{\frac{1}{4}+\varphi+\gamma}} \qquad (32)$$

Then, from the Rodrigue's equation, Eq. (6), the hypergeometric equation here is expressed in Jacobi polynomial as

$$y_n(s) = N_{nl} P_n^{\left(2\sqrt{\varepsilon_{nl}-\beta+\gamma},\, 2\sqrt{\frac{1}{4}+\varphi+\gamma}\right)}(1-2s) \qquad (33)$$

So that the wavefunction is

$$U_{nl}(s) = N_{nl} s^{\sqrt{\varepsilon_{nl}-\beta+\gamma}}(1-s)^{\frac{1}{2}+\sqrt{\frac{1}{4}+\varphi+\gamma}} P_n^{\left(2\sqrt{\varepsilon_{nl}-\beta+\gamma},\, 2\sqrt{\frac{1}{4}+\varphi+\gamma}\right)}(1-2s) \qquad (34)$$

From the definition of Jacobi polynomials [37],

$$P_n^{(\theta,\vartheta)}(\omega) = \frac{\Gamma(n+\theta+1)}{n!\,\Gamma(\theta+1)} {}_2F_1\left(-n,\theta+\vartheta+n+1,\theta+1;\frac{1-\omega}{2}\right) \qquad (35)$$

So, Eq. (34) is rewritten in terms of hypergeometric polynomial as

$$U_{nl}(s) = N_{nl} s^{A/2}(1-s)^{\frac{1}{2}+\sqrt{\frac{1}{4}+\varphi+\gamma}} \frac{\Gamma(n+A+1)}{n!\,\Gamma(2A+1)} {}_2F_1\left(-n, A+\sqrt{\frac{1}{4}+\varphi+\gamma}+n+1, A+1; s\right) \qquad (36)$$

Where $A = 2\sqrt{\varepsilon_{nl} - \beta + \gamma}$ (37)

From the normalization condition, we have that

$$\int_0^\infty |\psi(r)|^2 dr = 1 \tag{38}$$

And in our coordinate transformation as $s = e^{-\alpha r}$, we have that

$$-\frac{1}{\alpha s}\int_1^0 |U(s)|^2 ds = 1 \tag{39}$$

Now, letting $y = 1 - 2s$, we obtain

$$\frac{N_{n,l}(s)}{2\alpha}\int_{-1}^1 \left(\frac{1-y}{2}\right)^{A-1}\left(\frac{1+y}{2}\right)^{2\sqrt{\frac{1}{4}+\varphi+\gamma}+1}\left[P_n^{\left(A,2\sqrt{\frac{1}{4}+\varphi+\gamma}\right)}(y)\right]^2 dy = 1 \tag{40}$$

According to Onate et al,[38], integral of the form Eq. (40) can be expressed as

$$\int_{-1}^1 \left(\frac{1-p}{2}\right)^a\left(\frac{1+p}{2}\right)^b\left[P_n^{(a,b)}(p)\right]^2 dp = \frac{2\Gamma(a+n+1)\Gamma(b+n+1)}{n!a\Gamma(a+b+n+1)} \tag{41}$$

Therefore, comparing eq. (40) and (41), we obtain the normalization constant as

$$N_{n,l}(s) = \sqrt{\frac{n!\alpha(A-1)\Gamma\left(A+\sqrt{\frac{1}{4}+\varphi+\gamma}+n+1\right)}{\Gamma(\lambda+n)\Gamma\left(\sqrt{\frac{1}{4}+\varphi+\gamma}+n+2\right)}} \tag{42}$$

### 3.3 Fermionic massive spin 1/2 particles interacting with Hellmann-generalized Morse potential model

In this section, we briefly review the Dirac equation. The Dirac equation with scalar S(r) and V(r) potentials in spherical coordinates is given as [55]

$$[\vec{\alpha} \cdot \vec{p} + \beta(M + S(r)) - (E - V(r))]\psi(\vec{r}) = 0. \tag{43}$$

where E denotes the relativistic energy of the system, $p = -i\vec{\nabla}$ is the momentum operator and M is the mass of the fermionic particle. $\vec{\alpha}$ and $\beta$ are the 4×4 usual Dirac matrices given by

$$\vec{\alpha} = \begin{pmatrix} 0 & \vec{\sigma} \\ \vec{\sigma} & 0 \end{pmatrix}, \quad \beta = \begin{pmatrix} I & 0 \\ 0 & -I \end{pmatrix}, \tag{44}$$

where I is the 2×2 unitary matrix and $\vec{\sigma}$ are three-vector spin matrices

$$\sigma_1 = \begin{pmatrix} 0 & 1 \\ 1 & 0 \end{pmatrix}, \quad \sigma_2 = \begin{pmatrix} 0 & -i \\ i & 0 \end{pmatrix}, \quad \sigma_3 = \begin{pmatrix} 1 & 0 \\ 0 & -1 \end{pmatrix} \tag{45}$$

For a particle in a spherical field, the total angular momentum operator j and the spin-orbit matrix operator k = (σ. L + 1), where σ and L are the Pauli matrix and orbital angular momentum respectively, commute with the Dirac Hamiltonian. The eigenvalues of k are $k = -(j + 1/2)$ for the aligned spin $(s_{1/2}, p_{3/2}, \text{etc})$ and $k = (j + 1/2)$ for the unaligned spin $(p_{1/2}, d_{3/2}, \text{etc})$. The complete set of conservative quantities can be chosen as $(H, K, J^2, J_z)$. The Dirac spinor is

$$\psi_{n,\kappa}(\vec{r}) = \begin{pmatrix} f_{n,\kappa}(\vec{r}) \\ g_{n,\kappa}(\vec{r}) \end{pmatrix} = \begin{pmatrix} \frac{F_{n,\kappa}(r)}{r} Y_{jm}^{\ell}(\theta, \varphi) \\ \frac{G_{n,\kappa}(r)}{r} Y_{jm}^{\hat{\ell}}(\theta, \varphi) \end{pmatrix}, \tag{46}$$

where $F_{n,\kappa}(r)$ and $G_{n,\kappa}(r)$ are the radial wave functions of the upper and lower components respectively with $Y_{jm}^{\ell}(\theta, \varphi)$ and $Y_{jm}^{\hat{\ell}}(\theta, \varphi)$ for spin and pseudospin spherical harmonics coupled to the angular momentum on the z-axis. Now substitute Eq. (46) into Eq. (43), we recast the following differential equations [64]

$$\left(\frac{d^2}{dr^2} + \frac{\kappa}{r}\right) F_{n,\kappa}(r) = \left(M + E_{n\kappa} - V(r) + S(r)\right) G_{n,\kappa}(r), \tag{47}$$

$$\left(\frac{d^2}{dr^2} - \frac{\kappa}{r}\right) G_{n,\kappa}(r) = \left(M - E_{n\kappa} - V(r) - S(r)\right) F_{n,\kappa}(r) \tag{48}$$

which later translates to;

$$\left[\frac{d^2}{dr^2} - \frac{\kappa(\kappa+1)}{r^2} - \left[(M + E_{n\kappa} - \Delta(r))(M - E_{n\kappa} + \Sigma(r))\right] + \frac{\frac{d\Delta(r)}{dr}}{M + E_{n\kappa} - \Delta(r)}\left(\frac{d}{dr} + \frac{\kappa}{r}\right)\right] F_{n,\kappa}(r) = 0 \tag{49}$$

for $\kappa(\kappa + 1) = \ell(\ell + 1), r \in (0, \infty)$,

$$\left[\frac{d^2}{dr^2} - \frac{\kappa(\kappa-1)}{r^2} - \left[(M + E_{n\kappa} - \Delta(r))(M - E_{n\kappa} + \Sigma(r))\right] + \frac{\frac{d\Sigma(r)}{dr}}{M - E_{n\kappa} + \Sigma(r)}\left(\frac{d}{dr} - \frac{\kappa}{r}\right)\right] G_{n,\kappa}(r) = 0 \tag{50}$$

for $(\kappa - 1) = \hat{\ell}(\hat{\ell} + 1)$, $r \in (0, \infty)$, where $\Delta(r) = V(r) - S(r)$ and $\Sigma(r) = V(r) + S(r)$ are the difference and the sum potentials, respectively.

### 3.4 Spin symmetry solutions of the Dirac equation with Hellmann-generalized Morse potential model

Under this symmetry, $\frac{d[V(r)-S(r)]}{dr} = \frac{d\Delta(r)}{dr} = 0$, $\Delta(r) = C_s$ =constant, then the spin symmetry is exact in the Dirac equation. In this symmetry limit, we take $\Sigma(r)$ as the potential in the following form:

$$\Sigma(r) = -\frac{a}{r} + \frac{be^{-\alpha r}}{r} + D_e\left(1 - \frac{q}{e^{\alpha r}-1}\right)^2 \tag{51}$$

Substituting Eq. (51) into Eq. (49), we have;

$$\left[\frac{d^2}{dr^2} - \frac{\kappa(\kappa+1)}{r^2} - \left[(M + E_{n\kappa} - C_s)\left(M - E_{n\kappa} - \frac{a}{r} + \frac{be^{-\alpha r}}{r} + D_e\left(1 - \frac{q}{e^{\alpha r}-1}\right)^2\right)\right]\right] F_{n,\kappa}(r) = 0 \tag{52}$$

To obtain the solution for $\kappa \neq 0$, we use Eq.(15) [37] to deal with the centrifugal term, and transforming as $s = e^{-\alpha r}$ so as to make Eq. (52) conform to Eq. (2) satisfying the upper-spinor component $F_{n,\kappa}(s)$, then, we have

$$\frac{d^2 F_{n,\kappa}(s)}{ds^2} + \frac{(1-s)}{s(1-s)} \frac{dF_{n,\kappa}(s)}{ds} + \frac{1}{s^2(1-s)^2}[-(\delta_0 + \gamma_0 + \gamma_1 - \delta_2)s^2 + (2\gamma_0 + \delta_0 - \delta_1 - \delta_2)s - (\gamma_1 + \beta_1 - \delta_1)]F_{n,\kappa}(s) = 0 \tag{53}$$

Where

$$\beta_0 = M + E_{n\kappa} - C_s, \quad \beta_1 = \kappa(\kappa + 1), \quad \beta_2 = M - E_{n\kappa}, \quad \delta_0 = \frac{2\beta_0 D_e q}{\alpha^2}, \quad \delta_1 = \frac{\beta_0 a}{\alpha}, \quad \delta_2 = \frac{\beta_0 b}{\alpha}, \quad \gamma_0 = \frac{\beta_0 D_e q^2}{\alpha^2}, \quad \gamma_1 = \frac{\beta_0(\beta_2 + D_e)}{\alpha^2} \tag{54}$$

Comparing Eq. (53) and Eq. (2), as usual yields the following parameters

$$\begin{cases} \tilde{\tau}(s) = 1 - s \\ \sigma(s) = s(1-s) \\ \tilde{\sigma}(s) = -(\delta_0 + \gamma_0 + \gamma_1 - \delta_2)s^2 + (2\gamma_1 + \delta_0 - \delta_1 - \delta_2)s - (\gamma_1 + +\beta_1 - \delta_1) \end{cases} \tag{56}$$

Substituting these polynomials into Eq. (9), we get $\pi(s)$ to be

$$\pi(s) = -\frac{s}{2} \pm \sqrt{\left(\frac{1}{4} + (\delta_0 + \gamma_0 + \gamma_1 - \delta_2) - k\right)s^2 + \left(k - (2\gamma_1 + \delta_0 - \delta_1 - \delta_2)\right)s + \gamma_1 + \beta_1 - \delta_1} \tag{57}$$

Also, to find the constant $k$, the discriminant of the expression under the square root of Eq. (57) should be equal to zero. As such, we have that

$$k_{\pm} = -(\delta_0 + \delta_1 - \delta_2 - 2\beta_1) \pm 2\sqrt{\gamma_1 + +\beta_1 - \delta_1}\sqrt{\frac{1}{4} + \beta_1 + \gamma_0} \tag{58}$$

Substituting Eq. (58) into Eq. (58) yields

$$\pi(s) = -\frac{s}{2} \pm \left(\left(\sqrt{\gamma_1 + +\beta_1 - \delta_1} - \sqrt{\frac{1}{4} + \beta_1 + \gamma_0}\right)s - \sqrt{\gamma_1 + +\beta_1 - \delta_1}\right) \tag{59}$$

Taking the derivative of $\pi(s)$ in Eq.(59) with respect to s yields

$$\pi'_{-}(s) = -\frac{1}{2} - \left(\sqrt{\gamma_1 + +\beta_1 - \delta_1} - \sqrt{\frac{1}{4} + \beta_1 + \gamma_0}\right) \tag{60}$$

From the knowledge of NU method, we choose the expression $\pi(s)_{-}$ which the function $\tau(s)$ has a negative derivative. This is given by

$$k_{-} = -(\delta_0 + \delta_1 - \delta_2 - 2\beta_1) - 2\sqrt{\gamma_1 + +\beta_1 - \delta_1}\sqrt{\frac{1}{4} + \beta_1 + \gamma_0} \tag{61}$$

with $\tau(s)$ being obtained as

$$\tau(s) = 1 - 2s - 2\left(\sqrt{\gamma_1 + +\beta_1 - \delta_1} - \sqrt{\frac{1}{4} + \beta_1 + \gamma_0}\right)s + 2\sqrt{\gamma_1 + +\beta_1 - \delta_1} \tag{62}$$

Referring to Eq. (10), we define the constant $\lambda$ as

$$\lambda = -(\delta_0 + \delta_1 - \delta_2 - 2\beta_1) - 2\sqrt{\gamma_1 + +\beta_1 - \delta_1}\sqrt{\frac{1}{4} + \beta_1 + \gamma_0} - \frac{1}{2} - \left(\sqrt{\gamma_1 + +\beta_1 - \delta_1} - \sqrt{\frac{1}{4} + \beta_1 + \gamma_0}\right) \tag{63}$$

Taking the derivative of Eq. (62) with respect to s, we have;

$$\tau'(s) = -2 - 2\left(\sqrt{\gamma_1 + +\beta_1 - \delta_1} - \sqrt{\frac{1}{4} + \beta_1 + \gamma_0}\right) < 0 \tag{64}$$

From eq. (11), we obtain;

$$\lambda_n = n^2 + n + 2n\left(\sqrt{\frac{1}{4} + \beta_1 + \gamma_0} + \sqrt{\gamma_1 + +\beta_1 - \delta_1}\right) \tag{65}$$

By comparing Eqs. (63) and (65), the exact energy eigenvalue equation is obtained as

$$\gamma_1 = \delta_1 + \beta_1 + \frac{1}{4}\left[\frac{\left(n+\frac{1}{2}+\sqrt{\frac{1}{4}+\beta_1+\gamma_0}\right)^2 + \delta_2 - \delta_0 - \delta_1 + \beta_1 + \gamma_0}{\left(n+\frac{1}{2}+\sqrt{\frac{1}{4}+\beta_1+\gamma_0}\right)}\right]^2 \tag{66}$$

$$\beta_0 = M + E_{n\kappa} - C_s, \quad \beta_1 = \kappa(\kappa+1), \quad \beta_2 = M - E_{n\kappa}, \quad \delta_0 = \frac{2\beta_0 D_e q}{\alpha^2}, \quad \delta_1 = \frac{\beta_0 a}{\alpha}, \quad \delta_2 = \frac{\beta_0 b}{\alpha}, \quad \gamma_0 = \frac{\beta_0 D_e q^2}{\alpha^2}, \quad \gamma_1 = \frac{\beta_0(\beta_2 + D_e)}{\alpha^2} \tag{67}$$

$$(M + E_{n\kappa} - C_s)(M - E_{n\kappa} + D_e) = (M + E_{n\kappa} - C_s)a\alpha - \alpha^2\kappa(\kappa+1) + \frac{\alpha^2}{4}\left[\frac{\left(n+\frac{1}{2}+\sqrt{\frac{1}{4}+\kappa(\kappa+1)+(M+E_{n\kappa}-C_s)\frac{D_e q^2}{\alpha^2}}\right)^2 + (M+E_{n\kappa}-C_s)\left(\frac{b}{\alpha}-\frac{2D_e q}{\alpha^2}-\frac{a}{\alpha}+\frac{D_e q^2}{\alpha^2}\right) + \kappa(\kappa+1)}{\left(n+\frac{1}{2}+\sqrt{\frac{1}{4}+\kappa(\kappa+1)+(M+E_{n\kappa}-C_s)\frac{D_e q^2}{\alpha^2}}\right)}\right]^2 \tag{68}$$

The corresponding wave functions can be evaluated by substituting $\pi(s)_-$ and $\sigma(s)$ from Eq. (59) and Eq. (56) respectively into Eq. (5) and solving the first order differential equation, this gives the corresponding upper-spinor component as;

$$F_{n,\kappa}(s) = N_{n,\kappa} s^{\sqrt{\gamma_1+\beta_1-\delta_1}}(1-s)^{\frac{1}{2}+\sqrt{\frac{1}{4}+\beta_1+\gamma_0}} P_n^{\left(2\sqrt{\gamma_1+\beta_1-\delta_1},2\sqrt{\frac{1}{4}+\beta_1+\gamma_0}\right)}(1-2s) \tag{69}$$

From the definition of the Jacobi Polynomials [20],

$$P_n^{\left(2\omega,2\sqrt{\frac{1}{4}+\beta_1+\gamma_0}\right)}(1-2s) = \frac{(2\omega+1)_n}{n!}\,_2F_1\left(-n, 2\omega + 2\sqrt{\frac{1}{4}+\beta_1+\gamma_0} + n + 1, 2\omega + 1; s\right) \tag{70}$$

$$\omega = \sqrt{\gamma_1 + \beta_1 - \delta_1} \tag{71}$$

In terms of hypergeometric Polynomials, Eq. (69) can be written as

$$F_{n,\kappa}(s) = N_{n,\kappa}s^{\omega}(1-s)^{\frac{1}{2}+\sqrt{\frac{1}{4}+\beta_1+\gamma_0}} \frac{(2\omega+1)_n}{n!} {}_2F_1\left(-n, 1+2\omega + 2\sqrt{\frac{1}{4}+\beta_1+\gamma_0}+n; 2\omega+1; s\right) \tag{72}$$

where $(2\omega+1)_n$ is the Pochhammer's symbol(for the rising factorial).

Using the normalization condition, we obtain the normalization constant as follows:

$$\int_0^{\infty} F_{n,\kappa}(r) \times F_{n,\kappa}(r)^* dr = 1 \tag{73}$$

$$-\frac{1}{\alpha}\int_1^0 |F_{n,\kappa}(s)|^2 \frac{ds}{s} = 1, s = e^{-\alpha r} \tag{74}$$

$$\frac{1}{2\alpha}\int_{-1}^1 |F_{n,\kappa}(z)|^2 \frac{2}{1-z} dz = 1, z = 1 - 2s \tag{75}$$

Substituting Eq. (69) into Eq. (75), we have

$$\frac{N_{n,\kappa}^2}{2\alpha}\int_{-1}^1 \left(\frac{1-z}{2}\right)^{2\omega-1}\left(\frac{1+z}{2}\right)^{2\phi}\left[P_n^{(2\omega,2\phi-1)}(z)\right]^2 dz = 1, \tag{76}$$

where

$$\phi = \frac{1}{2} + \frac{1}{2}\sqrt{1+4(\beta_1+\gamma_0)}, \tag{77}$$

$$\omega = \sqrt{\gamma_1 + \beta_1 - \delta_1} \tag{78}$$

Comparing Eq. (76) with the integral of the form

$$\int_1^{-1}\left(\frac{1-p}{2}\right)^x\left(\frac{1+p}{2}\right)^y\left[P_n^{(2x,2y-1)}(p)\right]^2 dp = \frac{2\Gamma(x+n+1)\Gamma(y+n+1)}{n!x\Gamma(x+y+n+1)} \tag{79}$$

We have the normalization constant as

$$N_{n,\kappa} = \sqrt{\frac{n!2\omega\alpha\Gamma(2\omega+2\phi+n+1)}{\Gamma(2\omega+n+1)\Gamma(2\phi+n+1)}} \tag{80}$$

## 3.5 Pseudospin symmetry solutions of the Dirac equation with the Hellmann-generalized Morse potential model

In the pseudospin symmetry limit, $\frac{d\Sigma(r)}{dr} = 0$ and $\Sigma(r) = C_{ps} =$ constant. Thus, our potential is taken as

$$\Delta(r) = -\frac{a}{r} + \frac{be^{-\alpha r}}{r} + D_e\left(1 - \frac{q}{e^{\alpha r}-1}\right)^2 \tag{81}$$

substituting Eqs. (81) into Eq. (50), we obtain

$$\left[\frac{d^2}{dr^2} - \frac{\kappa(\kappa-1)}{r^2} - \left[\left(M + E_{n\kappa} + \frac{a}{r} - \frac{be^{-\alpha r}}{r} - D_e\left(1 - \frac{q}{e^{\alpha r}-1}\right)^2\right)(M - E_{n\kappa} + \Sigma(r))\right]\right] G_{n,\kappa}(r) = 0$$
(82)

Under this limit, similarly, we define $s = e^{-\alpha r}$, then, Eq. (82) becomes

$$\frac{d^2 G_{n,\kappa}(s)}{ds^2} + \frac{(1-s)}{s(1-s)}\frac{dG_{n,\kappa}(s)}{ds} + \frac{1}{s^2(1-s)^2}\left[-(\chi_0 + \chi_2 - \theta_1 - \theta_2)s^2 + (2\chi_0 + \chi_1 + \chi_2 - \theta_2)s - (\chi_0 + \lambda_1 + \chi_1)\right]G_{n,\kappa}(s) = 0$$
(83)

Where $\lambda_0 = M - E_{n\kappa} + C_{ps}$, $\lambda_1 = \kappa(\kappa - 1)$, $\lambda_2 = M + E_{n\kappa}$, $-\chi_0 = \frac{(D_e - \lambda_2)\lambda_0}{\alpha^2}$, $\chi_1 = \frac{\lambda_0 a}{\alpha}$, $\chi_2 = \frac{\lambda_0 b}{\alpha}$, $\theta_1 = \frac{D_e q^2 \lambda_0}{\alpha^2}$, $\theta_2 = \frac{2 D_e q \lambda_0}{\alpha^2}$
(84)

Using the same procedure as in spin symmetry above, we obtained the negative energy of the pseudospin symmetry as

$$(D_e - M - E_{n\kappa})(M - E_{n\kappa} + C_{ps}) = \kappa(\kappa-1)\alpha^2 + (M - E_{n\kappa} + C_{ps})a\alpha - \frac{\alpha^2}{4}\left[\frac{\left(n+\frac{1}{2}+\sqrt{\frac{1}{4}-\frac{D_e q^2(M-E_{n\kappa}+C_{ps})}{\alpha^2}+\kappa(\kappa-1)}\right)^2 + (M-E_{n\kappa}+C_{ps})\left(\frac{a}{\alpha}-\frac{b}{\alpha}+\frac{2D_e q}{\alpha^2}+\kappa(\kappa-1)+\frac{D_e q^2}{\alpha^2}\right)}{\left(n+\frac{1}{2}+\sqrt{\frac{1}{4}-\frac{D_e q^2(M-E_{n\kappa}+C_{ps})}{\alpha^2}+\kappa(\kappa-1)}\right)}\right]^2$$
(85)

The lower-spinor component of the wave function is obtained following the same procedure as;

$$G_{n,\kappa}(s) = N_{n,\kappa} s^{\sqrt{\chi_0+\lambda_1+\chi_1}}(1-s)^{\frac{1}{2}+\sqrt{\frac{1}{4}-\theta_1+\lambda_1}} P_n^{\left(2\sqrt{\chi_0+\lambda_1+\chi_1}, 2\sqrt{\frac{1}{4}-\theta_1+\lambda_1}\right)}(1-2s)$$
(86)

From the definition of the Jacobi Polynomials [20],

$$P_n^{\left(2\Omega, 2\sqrt{\frac{1}{4}-\theta_1+\lambda_1}\right)}(1-2s) = \frac{(2\Omega+1)_n}{n!} {}_2F_1\left(-n, 1+2\Omega + 2\sqrt{\frac{1}{4}-\theta_1+\lambda_1} + n; 2\Omega+1; s\right)$$
(87)

$$\Omega = \sqrt{\chi_0 + \lambda_1 + \chi_1}$$
(88)

In terms of hypergeometric Polynomials, Eq. (86) can be written as

$$G_{n,\kappa}(s) = N_{n,\kappa} s^{\omega}(1-s)^{\frac{1}{2}+\sqrt{\frac{1}{4}+\chi_1+\gamma}}\frac{(2\Omega+1)_n}{n!} {}_2F_1\left(-n, 1+2\Omega + 2\sqrt{\frac{1}{4}-\theta_1+\lambda_1} + n; 2\Omega+1; s\right)$$
(89)

where $(2\Omega + 1)_n$ is the Pochhammer's symbol(for the rising factorial).

### 3.6 Non-Relativistic Limit:

In this section, we obtained the non-relativistic limit of the spin symmetry limit. The nonrelativistic Schrödinger equation is bosonic in nature, i.e., spin does not involve in it. On the other hand, relativistic Dirac equation is for a spin 1/2 particle. This immediately suggests that there may be a certain relation between the solutions of the two fundamental equations [56-58]. The meeaning is that, the nonrelativistic Energies $E_{nl}$ can be determined by taking the non-relativistic limit values of the relativistic eigenenergies E. Therefore, taking $C_s = 0$ and using the transformations

$M + E_{n\kappa} \to \frac{2\mu}{\hbar^2}$ and $M - E_{n\kappa} \to -E_{n\ell}$ together with $\kappa \to \ell$ [56-57], the relativistic energy Eq. (68) reduces to

$$E_{n\ell} = D_e - a\alpha + \frac{\hbar^2\alpha^2\ell(\ell+1)}{2\mu} - \frac{\hbar^2\alpha^2}{8\mu}\left[\frac{\left(n+\frac{1}{2}+\sqrt{\frac{1}{4}+\ell(\ell+1)+\frac{2\mu D_e q^2}{\hbar^2\alpha^2}}\right)^2 + \frac{2\mu}{\hbar^2}\left(\frac{b}{\alpha}-\frac{2D_e q}{\alpha^2}-\frac{a}{\alpha}+\frac{D_e q^2}{\alpha^2}\right)+\ell(\ell+1)}{\left(n+\frac{1}{2}+\sqrt{\frac{1}{4}+\ell(\ell+1)+\frac{2\mu D_e q^2}{\hbar^2\alpha^2}}\right)}\right]^2 \quad (90)$$

$$F_{n,\ell}(s) = N_{n,\ell} s^\omega (1-s)^\phi \frac{(2\omega+1)_n}{n!} {}_2F_1(-n, 2\omega+\phi+n; 2\omega+1; s) \quad (91)$$

We have the normalization constant as

$$N_{n,\ell} = \sqrt{\frac{n! 2\omega\alpha \Gamma(2\omega+2\phi+n+1)}{\Gamma(2\omega+n+1)\Gamma(2\phi+n+1)}} \quad (92)$$

$$\phi = \frac{1}{2} + \sqrt{\frac{1}{4} + \ell(\ell+1) + \frac{2\mu D_e q^2}{\hbar^2\alpha^2}}, \quad (93)$$

$$\omega = \sqrt{\frac{2\mu(D_e - E_{nl})}{\hbar^2\alpha^2} + \ell(\ell+1) - \frac{2\mu a}{\hbar^2\alpha}} \quad (94)$$

### 4. Numerical results and discussion

By using the well-known spectroscopic values in Table 1, we computed the energy eigenvalues of the Hellmann- generalized Morse Potential for some selected diatomic molecules (for $N_2$, CO, NO, CH and HCl) using the Non-Relativistic energy equation for various vibrational n and rotational $\ell$ quantum number, as shown in Tables 2. In Fig. 1, we plot the shape of the potential. This shows explicitly the behaviour of the potentil under consideration.In Fig. 2, we show the variation of the energy spectrum with various values of the rotational and vibrational quantum numbers. It is seen that for fixed value of $\ell$, the energy eigenvalues increases as n increases. There's also a spread at n = 0 but a uniform convergence as n $\to$ 10. In Figure 3 and 4, we show the variation of the Energy eigenvalues with dissociation energy (in $cm^{-1}$) and equillibrium bond length r (in Å) respectively for various vibrational quantum number. The energy increases monotonically in both cases. We observe in Fig. 5 that as the adjustable screening parameter increase, the energy eigenvalue increases in a quasi-linear manner. There is a uniform decrease in the energy eigenvalues as the parameter "a" increases, this is shown in Fig. 6. The energy eigenvalue increases as the parameter "b" increases up to b = 3, beyond this point, the energy decreases.

### 5 Conclusion

In this paper, approximate bound-state solutions of the Schrödinger, Klein–Gordon and Dirac equations for the Hellmann- generalized Morse model potential was considered. For each of the wave equation, and by applying an appropriate approximation to the centrifugal term, we have obtained the energy eigenvalues and the corresponding wave functions for any quantum state. To show the accuracy of our findings, we compute the numerical energy spectra for some diatomic molecules using some spectroscopic parameters and figures that discuss the energy spectrum in each case. It is interesting to note here that the results obtained from this consideration finds application in various branches of physics and chemistry where non-relativistic and relativistic phenomena are studied.

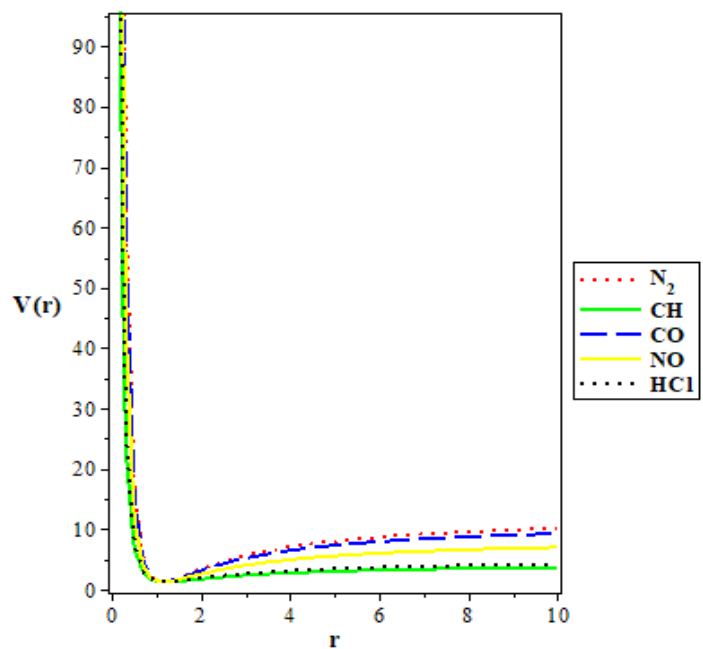

Fig. 1 The shape of Hellmann Generalised Morse potential (r in amstrong) for the slected diatomic molecules

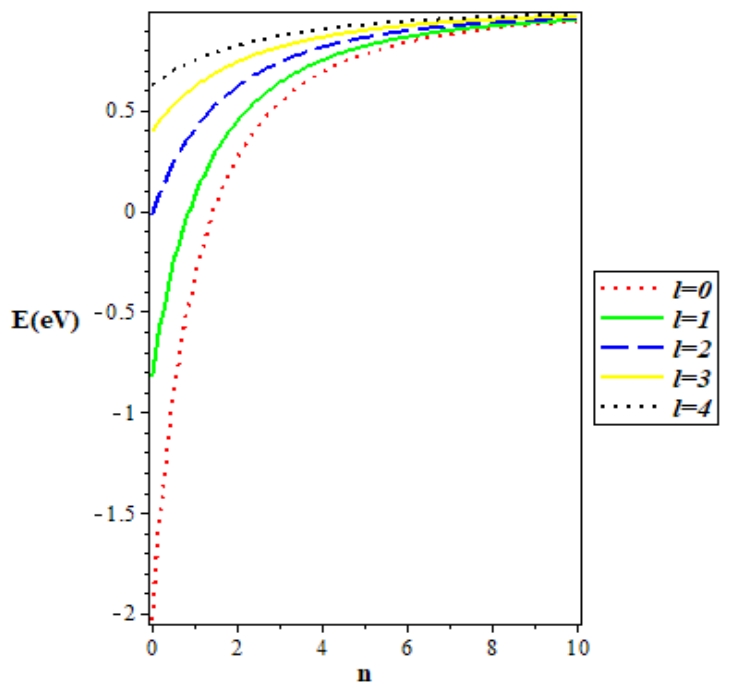

Fig. 2 Energy eigenvalues variation with various rotational and vibrational quantum numbers

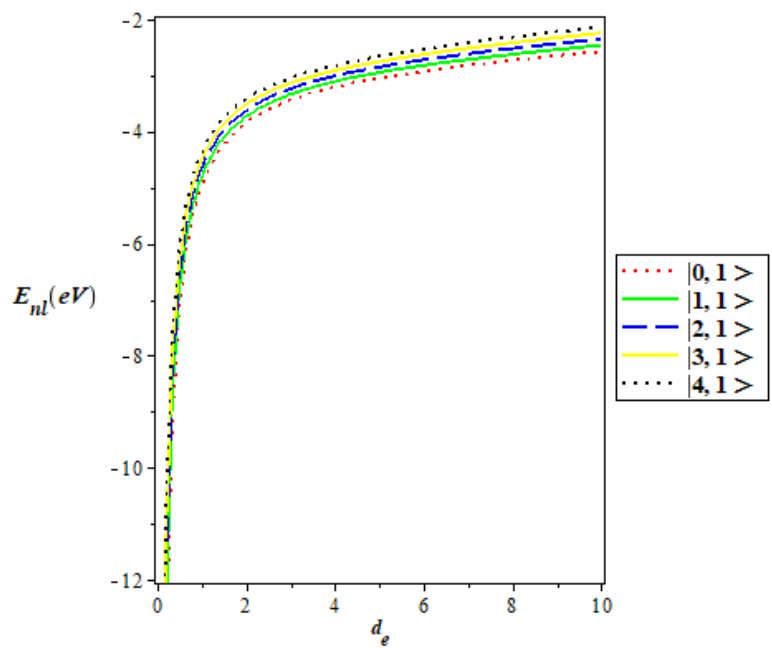

Fig. 3 Energy eigenvalues variation with dissociation energy (in $cm^{-1}$) for various vibrational quantum number

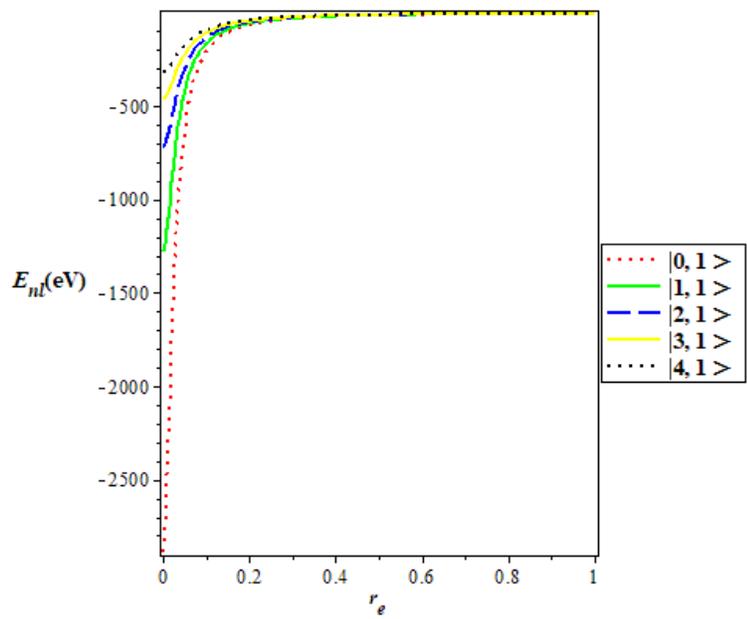

Fig. 4 Energy eigenvalues variation with equillibrium bond length r (in Å) for various vibrational quantum number

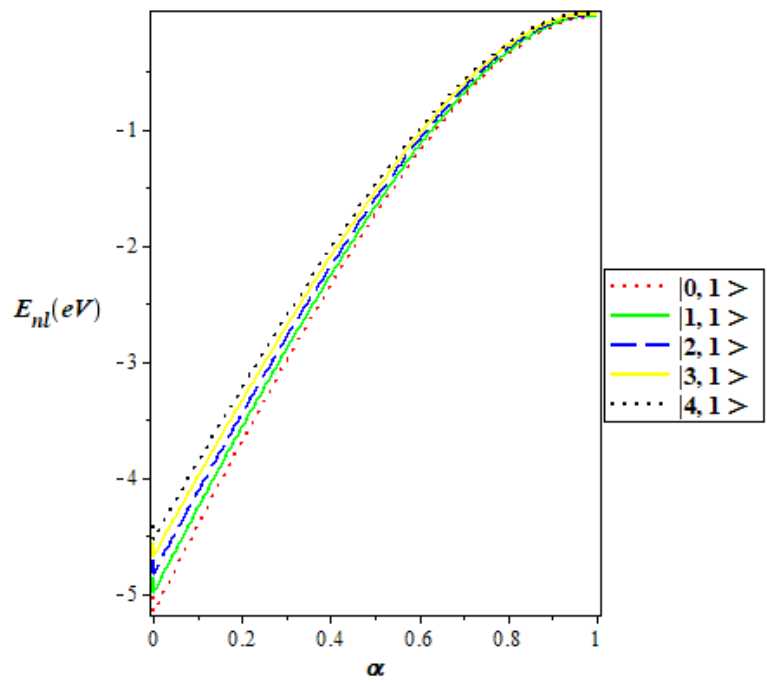

Fig. 5 Energy eigenvalues variation with adjustable screening parmeter for various vibrational quantum number

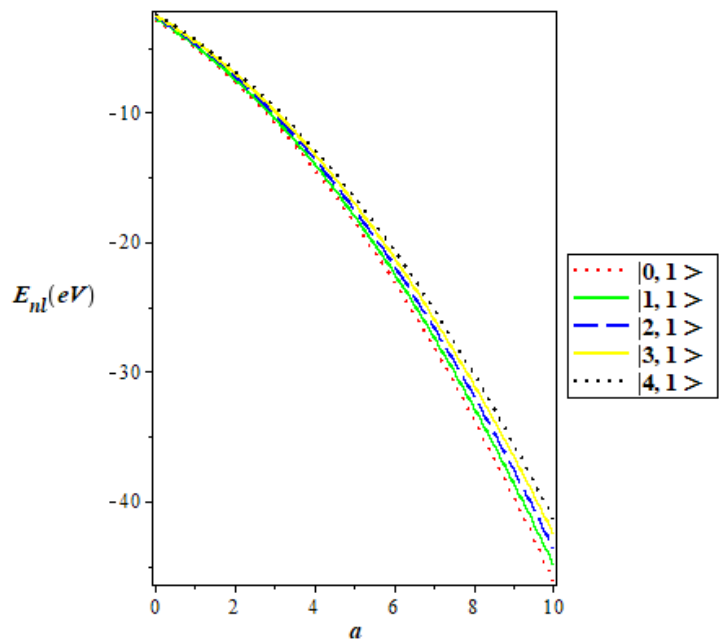

Fig. 6 Energy eigenvalues variation with potential strength for various vibrational quantum number

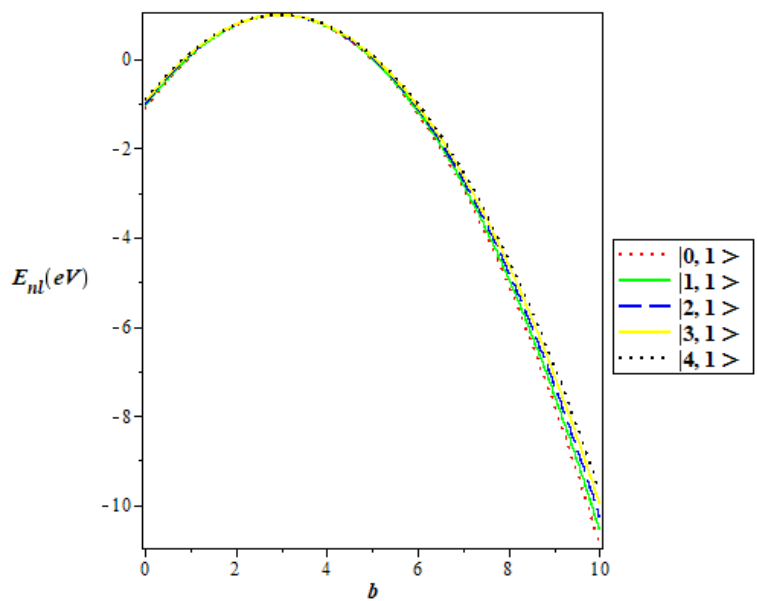

Fig. 7 Energy eigenvalues variation with coupling strength for various vibrational quantum number

Table 1; The potential model parameters for some diatomic molecules[59]

| Molecules | $D_e(cm^{-1})$ | $r_e(Å)$ | $\mu(amu)$ |
|---|---|---|---|
| CH | 31838.08 | 1.1198 | 0.929931 |
| NO | 64877.06 | 1.1508 | 7.468441 |
| CO | 87471.43 | 1.1282 | 6.860586 |
| $N_2$ | 96288.04 | 1.0940 | 7.003350 |
| HCl | 37,255.00 | 1.2746 | 0.980105 |

Table 2 Energy eigenvalues (in eV) in 3D for various vibrational quantum number and $\alpha = 0.025$ for CH, NO, CO $N_2$, and HCl molecules with $\hbar c = 1973.29$ eVÅ

| n | l | CH | NO | CO | $N_2$ | HCl |
|---|---|---|---|---|---|---|
| 0 | 0 | -2.669578479 | -2.208833696 | -2.057168210 | -2.090263840 | -2.146705710 |
| 1 | 0 | -2.397864979 | -2.103866806 | -1.935855630 | -1.961918940 | -1.924396284 |
|   | 1 | -2.392210914 | -2.103326846 | -1.935282930 | -1.961329540 | -1.920701408 |
| 2 | 0 | -2.142262376 | -2.000467316 | -1.816205960 | -1.835286920 | -1.712628335 |
|   | 1 | -2.136937738 | -1.999935346 | -1.815641030 | -1.834705280 | -1.709105695 |
|   | 2 | -2.126315533 | -1.998871537 | -1.814511320 | -1.833542130 | -1.702071683 |
| 3 | 0 | -1.901521961 | -1.898604212 | -1.698188930 | -1.710337410 | -1.510745969 |
|   | 1 | -1.896501566 | -1.898080047 | -1.697631630 | -1.709763420 | -1.507384983 |
|   | 2 | -1.886485915 | -1.897031875 | -1.696517160 | -1.708615560 | -1.500673639 |
|   | 3 | -1.871524996 | -1.895460006 | -1.694845800 | -1.706894110 | -1.490633087 |
| 4 | 0 | -1.674513710 | -1.798247185 | -1.581774980 | -1.587040750 | -1.318143535 |
|   | 1 | -1.669774726 | -1.797730694 | -1.581225160 | -1.586474270 | -1.314934415 |
|   | 2 | -1.660320149 | -1.796697836 | -1.580125670 | -1.585341430 | -1.308526216 |
|   | 3 | -1.646196510 | -1.795148948 | -1.578476760 | -1.583642500 | -1.298938890 |
|   | 4 | -1.627472860 | -1.793084493 | -1.576278860 | -1.581377920 | -1.286202222 |
| 5 | 0 | -1.460212987 | -1.699366720 | -1.466935230 | -1.465367930 | -1.134261051 |
|   | 1 | -1.455734691 | -1.698857735 | -1.466392760 | -1.464808840 | -1.131194757 |
|   | 2 | -1.446799912 | -1.697839905 | -1.465307970 | -1.463690750 | -1.125071629 |
|   | 3 | -1.433452000 | -1.696313552 | -1.463681110 | -1.462013950 | -1.115910550 |
|   | 4 | -1.415755311 | -1.694279120 | -1.461512590 | -1.459778880 | -1.103739628 |
|   | 5 | -1.393794428 | -1.691737223 | -1.458803000 | -1.456986010 | -1.088596024 |